\documentclass[aps,prd,preprint,nofootinbib,superscriptaddress,showpacs,byrevtex,titlepage,tightenlines]{revtex4}

\usepackage{graphicx} 
\usepackage{epsfig}
\usepackage{dcolumn}  

\graphicspath{{ps}}

\begin{document}
\preprint{\vbox{ \vskip 5mm
                 \hbox{\hskip-160mm
                       \includegraphics[height=2.cm]{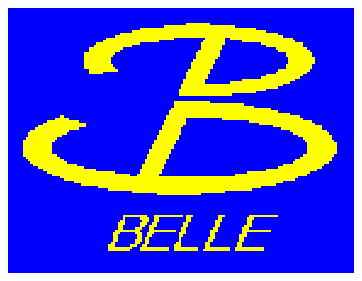}
		\hskip 120mm} \vskip -15mm
	         \hbox{KEK Preprint 2003-8}
                 \hbox{Belle Preprint 2003-3}
                 \hbox{hep-ex/0304032}
}}

\title{
\quad\\[1cm] \Large {\boldmath{Measurement of branching fraction ratios and CP asymmetries in $B^{\pm} \rightarrow D_{CP}K^{\pm}$}}}

\affiliation{Aomori University, Aomori}
\affiliation{Budker Institute of Nuclear Physics, Novosibirsk}
\affiliation{Chiba University, Chiba}
\affiliation{Chuo University, Tokyo}
\affiliation{University of Cincinnati, Cincinnati, Ohio 45221}
\affiliation{University of Frankfurt, Frankfurt}
\affiliation{Gyeongsang National University, Chinju}
\affiliation{University of Hawaii, Honolulu, Hawaii 96822}
\affiliation{High Energy Accelerator Research Organization (KEK), Tsukuba}
\affiliation{Hiroshima Institute of Technology, Hiroshima}
\affiliation{Institute of High Energy Physics, Chinese Academy of Sciences, Beijing}
\affiliation{Institute of High Energy Physics, Vienna}
\affiliation{Institute for Theoretical and Experimental Physics, Moscow}
\affiliation{J. Stefan Institute, Ljubljana}
\affiliation{Kanagawa University, Yokohama}
\affiliation{Korea University, Seoul}
\affiliation{Kyoto University, Kyoto}
\affiliation{Kyungpook National University, Taegu}
\affiliation{Institut de Physique des Hautes \'Energies, Universit\'e de Lausanne, Lausanne}
\affiliation{University of Ljubljana, Ljubljana}
\affiliation{University of Maribor, Maribor}
\affiliation{University of Melbourne, Victoria}
\affiliation{Nagoya University, Nagoya}
\affiliation{Nara Women's University, Nara}
\affiliation{National Kaohsiung Normal University, Kaohsiung}
\affiliation{National Lien-Ho Institute of Technology, Miao Li}
\affiliation{National Taiwan University, Taipei}
\affiliation{H. Niewodniczanski Institute of Nuclear Physics, Krakow}
\affiliation{Nihon Dental College, Niigata}
\affiliation{Niigata University, Niigata}
\affiliation{Osaka City University, Osaka}
\affiliation{Osaka University, Osaka}
\affiliation{Panjab University, Chandigarh}
\affiliation{Peking University, Beijing}
\affiliation{Princeton University, Princeton, New Jersey 08545}
\affiliation{RIKEN BNL Research Center, Upton, New York 11973}
\affiliation{Saga University, Saga}
\affiliation{University of Science and Technology of China, Hefei}
\affiliation{Seoul National University, Seoul}
\affiliation{Sungkyunkwan University, Suwon}
\affiliation{University of Sydney, Sydney NSW}
\affiliation{Tata Institute of Fundamental Research, Bombay}
\affiliation{Toho University, Funabashi}
\affiliation{Tohoku Gakuin University, Tagajo}
\affiliation{Tohoku University, Sendai}
\affiliation{Department of Physics, University of Tokyo, Tokyo}
\affiliation{Tokyo Institute of Technology, Tokyo}
\affiliation{Tokyo Metropolitan University, Tokyo}
\affiliation{Tokyo University of Agriculture and Technology, Tokyo}
\affiliation{Toyama National College of Maritime Technology, Toyama}
\affiliation{University of Tsukuba, Tsukuba}
\affiliation{Utkal University, Bhubaneswer}
\affiliation{Virginia Polytechnic Institute and State University, Blacksburg, Virginia 24061}
\affiliation{Yokkaichi University, Yokkaichi}
\affiliation{Yonsei University, Seoul}
  \author{S.~K.~Swain}\affiliation{University of Hawaii, Honolulu, Hawaii 96822} 
  \author{T.~E.~Browder}\affiliation{University of Hawaii, Honolulu, Hawaii 96822} 
  \author{K.~Abe}\affiliation{High Energy Accelerator Research Organization (KEK), Tsukuba} 
  \author{T.~Abe}\affiliation{Tohoku University, Sendai} 
  \author{I.~Adachi}\affiliation{High Energy Accelerator Research Organization (KEK), Tsukuba} 
  \author{H.~Aihara}\affiliation{Department of Physics, University of Tokyo, Tokyo} 
  \author{M.~Akatsu}\affiliation{Nagoya University, Nagoya} 
  \author{Y.~Asano}\affiliation{University of Tsukuba, Tsukuba} 
  \author{T.~Aso}\affiliation{Toyama National College of Maritime Technology, Toyama} 
  \author{T.~Aushev}\affiliation{Institute for Theoretical and Experimental Physics, Moscow} 
% ...
\collaboration{The Belle Collaboration}

\begin{abstract}
\noindent
	
We report results on the decay $B^{-} \rightarrow D_{CP}K^{-}$ and
its charge conjugate using a data sample of  85.4 million $B\overline{B}$
pairs recorded at the  $\Upsilon(4S)$ resonance with the Belle detector at the KEKB asymmetric $e^{+}e^{-}$ storage ring. Ratios of
branching fractions of Cabibbo-suppressed to Cabibbo-favored processes
are determined to be ${\cal B}(B^- \rightarrow D^0 K^-)/{\cal B}(B^-
\rightarrow D^0 \pi^-)= 0.077 \pm 0.005(stat) \pm 0.006(sys)$, ${\cal
B}(B^- \rightarrow D_1 K^-)/{\cal B}(B^- \rightarrow D_1 \pi^-) =
0.093 \pm 0.018(stat) \pm 0.008(sys)$ and ${\cal B}(B^- \rightarrow
D_2 K^-)/{\cal B}(B^- \rightarrow D_2 \pi^-) = 0.108 \pm 0.019(stat)
\pm 0.007(sys)$ where the indices 1 and 2 represent the CP=+1 and
CP=$-$1 eigenstates of the $D^{0}-\bar{D^{0}}$ system,
respectively. We find the partial-rate charge asymmetries for  $B^{-} \rightarrow D_{CP}K^{-}$ to be ${\cal{A}}_1 = 0.06 \pm 0.19(stat) \pm 0.04(sys)$ and ${\cal{A}}_2 = -0.19 \pm 0.17(stat) \pm 0.05(sys)$.
\end{abstract}
\maketitle \tighten
{\renewcommand{\thefootnote}{\fnsymbol{footnote}}}
\setcounter{footnote}{0}
\normalsize

The extraction of $\phi_{3}$~\cite{gamma}, an angle of the Kobayashi-Maskawa triangle~\cite{KM}, is a challenging measurement even with modern high luminosity $B$ factories. Recent theoretical work on $B$ meson dynamics has demonstrated the direct accessibility of $\phi_{3}$ using the process $B^{-} \rightarrow DK^{-}$~\cite{ADS,Gronau}. If the $D^{0}$ is reconstructed as a CP eigenstate, the $b \rightarrow c$ and $b \rightarrow u$ processes interfere. This interference leads to direct CP violation as well as a characteristic pattern of branching fractions. However, the branching fractions for $D$ meson decay modes to CP eigenstates are only of order 1~\%. Since CP violation through interference is expected to be small, a large number of $B$ decays is needed to extract $\phi_3$.  Assuming the absence of $D^0 - \bar{D^0}$ mixing, the observables sensitive to CP violation that are used to extract the angle $\phi_3$~\cite{BABAR} are,
\begin{eqnarray*}
{\cal{A}}_{1,2} \equiv \frac{{\cal B}(B^- \rightarrow D_{1,2}K^-) -
{\cal B}(B^+ \rightarrow D_{1,2}K^+) }{{\cal B}(B^- \rightarrow D_{1,2}K^-) + {\cal B}(B^+ \rightarrow D_{1,2}K^+) }&\\
= \frac{2 r \sin \delta ' \sin \phi_3}{1 + r^2 + 2 r \cos \delta ' \cos \phi_3}~~~~~~~~~~~~~~~~~~~~~~&\\
{\cal{R}}_{1,2} \equiv \frac{R^{D_{1,2}}}{R^{D^{0}}}  = 1 + r^2 + 2 r \cos \delta ' \cos \phi_3~~~~~~~~~~& \\
\delta ' = \left\{
             \begin{array}{ll}
              \delta & \mbox {{\rm  for }$D_1$}\\
              \delta + \pi&  \mbox{{\rm for }$D_2$}\\
             \end{array}
             \right.,~~~~~~~~~~~~~~~~~&
\end{eqnarray*}
where the ratios $R^{D_{1,2}}$ and $R^{D^{0}}$ are defined as
$$R^{D_{1,2}}=\frac{{\cal B}(B^- \rightarrow D_{1,2}K^-)+{\cal B}(B^+
\rightarrow D_{1,2}K^+)}{{\cal B}(B^- \rightarrow D_{1,2}\pi^-) +
{\cal B}(B^+ \rightarrow D_{1,2}\pi^+)},$$$$R^{D^{0}}=\frac{{\cal
B}(B^- \rightarrow D^{0} K^-)+{\cal B}(B^+ \rightarrow
\bar{D^{0}} K^+)}{{\cal B}(B^- \rightarrow D^{0}\pi^-) + {\cal B}(B^+
\rightarrow \bar{D^{0}} \pi^+)},$$$D_1$ and $D_2$ are CP-even and
CP-odd eigenstates of the neutral $D$ meson, $r = |A(B^- \to
\bar{D^0}K^-)/A(B^- \to D^0 K^-)|$ is the ratio of the amplitudes of
the two tree diagrams shown in Fig.~1 and $\delta$ is their strong-phase difference. The ratio $r$ corresponds to the magnitude of CP asymmetry and is suppressed to the level of $\sim 0.1$ due to the CKM factor~$(\sim 0.4)$ and a color suppression factor~$(\sim 0.25)$. Note that the asymmetries ${\cal{A}}_{1}$ and ${\cal{A}}_{2}$ have opposite signs. 

\begin{figure}[ht]
\begin{center}
  \epsfig{file=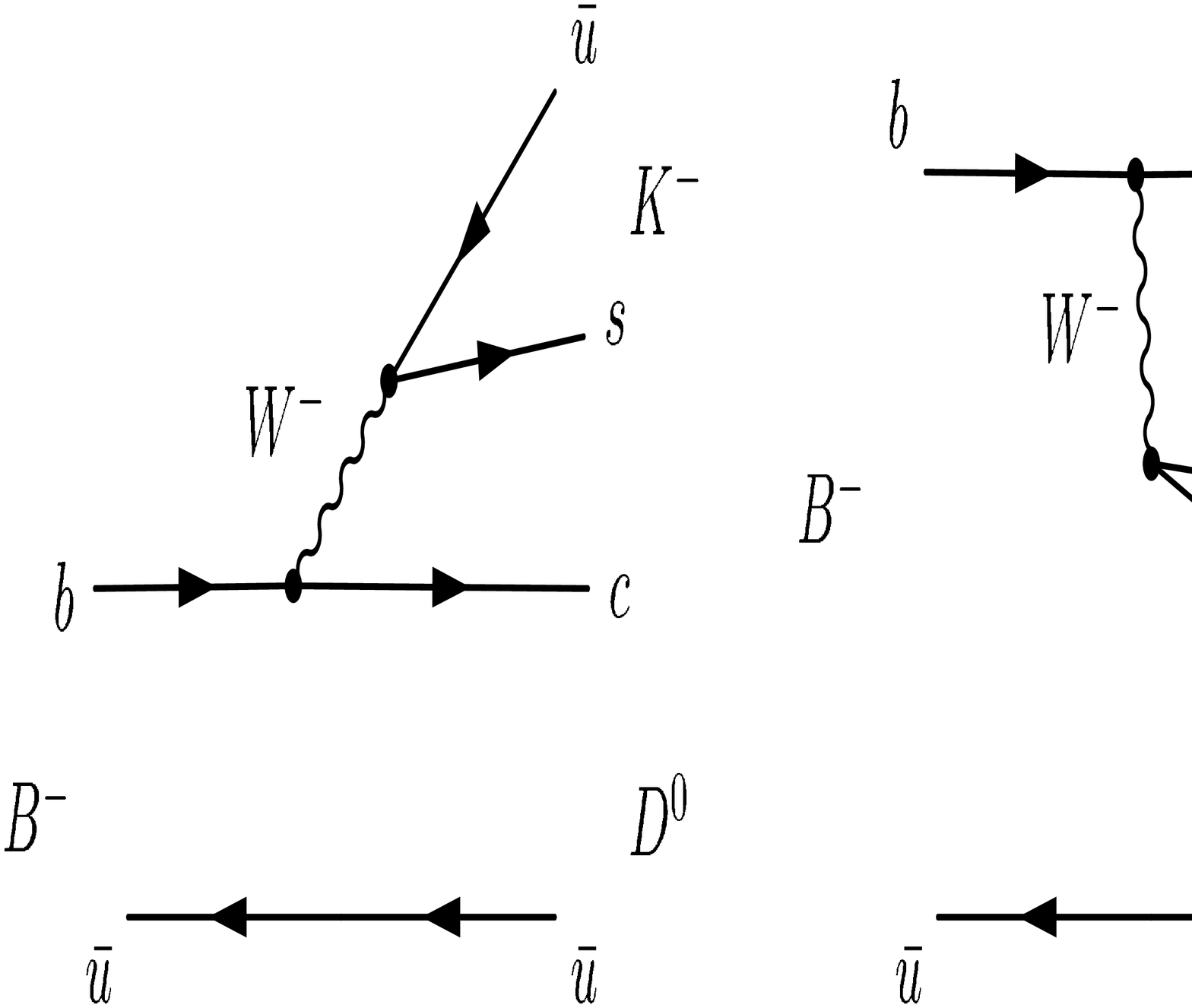,height=3.0cm,width=14cm} 
\end{center}
\caption{$B^{-} \rightarrow D^{0}K^{-}$ and $B^{-} \rightarrow \bar{D^{0}}K^{-}$.}
\end{figure}
The ratio of the Cabibbo-suppressed decay $B^{-} \rightarrow
D^{0}K^{-}$ to the Cabibbo-favored decay $B^{-} \rightarrow
D^{0}\pi^{-}$ has been reported by CLEO~\cite{CLEO} to be
$R^{D^{0}} = 0.099^{+0.014}_{-0.012}\hskip 1mm
^{+0.007}_{-0.006}$ while Belle finds $R^{D^0} = 0.079 \pm 0.009 \pm 0.006$~\cite{belledk}. Assuming factorization, the ratio $R^{D^{0}}$ is expected to be $\tan^2\theta_C(f_K/f_\pi)^2 \approx 0.074$ in the tree-level approximation, where $\theta_C$ is the Cabibbo angle, and $f_K$ and $f_\pi$ are meson decay constants. The measurements are in good agreement with this theoretical expectation.

Previously, Belle reported the observation of the decays $B^{-}
\rightarrow D_{1}K^{-}$ and $B^{-} \rightarrow D_{2}K^{-}$ with
$29.1~{\rm fb}^{-1}$~\cite{sugi}. This paper reports more precise
measurements of these decays with a data sample of $78~{\rm fb}^{-1}$,
containing 85.4 million $B\overline{B}$ pairs, 
collected  with the Belle detector at the KEKB asymmetric-energy
$e^+e^-$ (3.5 on 8~GeV) collider operating at the $\Upsilon(4S)$ resonance.
At KEKB, the $\Upsilon(4S)$ is produced with a Lorentz boost of
$\beta\gamma=0.425$ nearly along the electron beamline.

The Belle detector is a large-solid-angle magnetic
spectrometer that consists of a three-layer Silicon vertex detector (SVD),
a 50-layer central drift chamber (CDC), an array of
Silica aerogel threshold \v{C}erenkov counters (ACC), 
a barrel-like arrangement of time-of-flight
scintillation counters (TOF), and an electromagnetic calorimeter (ECL)
comprised of CsI(Tl) crystals located inside 
a super-conducting solenoid coil that provides a 1.5~T
magnetic field.  An iron flux-return located outside of
the coil is instrumented to detect $K_L^0$ mesons and to identify
muons (KLM).  The detector
is described in detail elsewhere~\cite{Belle}.

 We reconstruct $D^{0}$ mesons in the following decay channels. For
the flavor specific mode (denoted by $D_{f}$), we use $D^{0}
\rightarrow K^{-}\pi^{+}$~\cite{CC}. For CP =+1 modes, we use $D_{1}
\rightarrow K^{-}K^{+}$ and $\pi^{-}\pi^{+}$ while for CP =$-$1 modes,
we use $D_{2} \rightarrow K_{S}^{0}\pi^{0}$, $K_{S}^{0}\phi$,
$K_{S}^{0}\omega$, $K_{S}^{0}\eta$ and $K_{S}^{0}\eta'$. 

The charged track, $K_{S}^{0}$ and $\pi^{0}$ selection requirements
have been described in Ref.~\cite{sugi}. For each charged track, information from the ACC, TOF and specific ionization measurements
from the CDC are used to determine a $K/\pi$ likelihood ratio
$P(K/\pi)$ = $\it L_{K}/(\it L_{K} + \it L_{\pi})$, where $\it L_{K}$
and $\it L_{\pi}$ are kaon and pion likelihoods. For kaons (pions)
from the $D^{0} \rightarrow K^{-}\pi^{+}$ mode we used the particle
identification requirement of $P(K/\pi) > 0.4~(<0.7)$. For kaons from
the $D^{0} \rightarrow K^{-}K^{+}$ mode we require $P(K/\pi) > 0.7$
while for pions from $D^{0} \rightarrow \pi^{-}\pi^{+}$ mode we require $P(K/\pi) < 0.7$. 

The $\omega$ mesons are reconstructed from $\pi^{+}\pi^{-}\pi^{0}$
combinations in the mass window $0.732\,\mathrm{~GeV}/c^2 <
M(\pi^{+}\pi^{-}\pi^{0}) < 0.82\,\mathrm{~GeV}/c^2$ with the charged pion particle identification requirement
$P(K/\pi) < 0.8$. To reduce the contribution from the non-resonant
background, a helicity angle cut
$|{\rm{cos}}~\theta_{\rm{hel}}| > 0.4$ is applied where $\theta_{\rm{hel}}$
is the angle between the normal to the $\omega$ decay plane in the
$\omega$ rest frame and the $\omega$ momentum in the $D^{0}$ rest
frame. To remove the contribution from $D^{0} \rightarrow
K^{*-}\rho^{+}$, we require the $K_{S}^{0}\pi^{-}$ invariant mass to
be greater than $75\,\mathrm{~MeV}/c^2$ from the $K^{*-}$ nominal mass. 

The $\phi$ mesons are reconstructed from two oppositely charged kaons
in the mass window of $1.008\,\mathrm{~GeV}/c^2 <
M(K^{+}K^{-}) < 1.032\,\mathrm{~GeV}/c^2$ with $P(K/\pi) > 0.2$. We also apply the $\phi$
helicity angle cut $|{\rm{cos}}~\theta_{\rm{hel}}| > 0.4$ where
$\theta_{\rm{hel}}$ is the angle between one of the $\phi$ daughters in
the $\phi$ rest frame and the $\phi$ momentum in the $D^{0}$ rest
frame. We form candidate $\eta$ and $\eta'$ mesons using the
$\gamma\gamma$ and $\eta\pi^{+}\pi^{-}$ decay modes with mass ranges
of  $0.495\,\mathrm{~GeV}/c^2 < M(\gamma\gamma) <
0.578\,\mathrm{~GeV}/c^2$ and  $0.903\,\mathrm{~GeV}/c^2 <
M(\eta\pi^{+}\pi^{-}) < 1.002\,\mathrm{~GeV}/c^2$, respectively. The
$\eta$ momentum is required to be greater than $0.5\,\mathrm{~GeV}/c$. Both
$\eta$ and $\eta'$ candidates are kinematically constrained to
their nominal masses. The $D^{0}$ candidates are required to have
masses within $\pm$2.5$\sigma$ of their nominal masses, where $\sigma$
is the measured mass resolution which ranges from $4.9\,\mathrm{~MeV}$ to
$17.7\,\mathrm{~MeV}$ depending on the decay channel. A $D^{0}$ mass and (wherever possible)vertex constrained fit is then performed on the remaining candidates. 
   
We combine the $D^{0}$ and $\pi^{-}$/$K^{-}$ candidates (denoted by
$h$) to form $B$ candidates. We apply tighter particle identification
cuts, $P(K/\pi) > 0.8 (<0.8)$ for prompt kaons (pions), to identify
$B^{-} \rightarrow D^{0}K^{-}(\pi^{-})$ events. The signal is
identified by two kinematic variables calculated in the center-of-mass
(c.m.) frame. The first is the beam-energy constrained mass,
$M_{\rm{bc}} = \sqrt{E_{\rm{beam}}^2 - |\vec{p}_D + \vec{p}_{h}|^2}$,
where $\vec{p}_{D}$ and $\vec{p}_{h}$ are the momenta of $D^{0}$ and
$K^-/\pi^{-}$ candidates and $E_{\rm{beam}}$ is the beam energy in the
c.m. frame. The second is the energy difference, $\Delta E = E_D +
E_{h} - E_{\rm{beam}}$, where $E_D$ is the energy of the $D^{0}$
candidate, $E_{h}$ is the energy of the $K^-/\pi^{-}$ candidate
calculated from the measured momentum and assuming the pion
mass,$~E_{h}=\sqrt{|\vec{p}_{h}|^2+ m_{\pi}^2}$. With this
definition, real $B^{-} \rightarrow D^{0}\pi^{-}$ events peak at
$\Delta E=0$ even when they are misidentified as $B^{-} \rightarrow
D^{0}K^{-}$, while $B^{-} \rightarrow D^{0}K^{-}$ events peak around
$\Delta E=-49$ MeV. Event candidates are accepted if they have
$5.2\,\mathrm{~GeV}/c^2 < M_{\rm{bc}} < 5.3\,\mathrm{~GeV}/c^2$ and
$|\Delta E| < 0.2\,\mathrm{~GeV}$. In case of multiple candidates from
a single event, we choose the best candidate on the basis of a $\chi^2$ determined from the differences between the measured and nominal values of $M_D$ and $M_{\rm{bc}}$. 

To suppress the large combinatorial background from the two-jet like
$e^{+}e^{-} \rightarrow q\bar{q}$ ($q = u$, $d$, $s$ or $c$) continuum
processes, variables that characterize the event topology are used. We
construct a Fisher discriminant $\it F$, from 6 modified Fox Wolfram
moments~\cite{SFW}. Furthermore, $\cos{\theta_{B}}$, the angle of the
$B$ flight direction with respect to the beam axis is also used to
distinguish signal from continuum background. We combine these two
independent variables, $\it F$ and $\cos{\theta_{B}}$ to make a single
likelihood ratio variable ($LR$) that distinguishes signal from
continuum background. We apply a different requirement for each
sub-mode based on the expected signal yield and the backgrounds in the
$M_{\rm{bc}}$ sideband data. For $B^{-} \rightarrow D^{0}\pi^{-}$
where $D^{0} \rightarrow K^-\pi^+$, $K^-K^+$ we require $LR > 0.4$
whereas for $D^{0} \rightarrow \pi^{+}\pi^{-}$, $K_{S}^{0}\pi^{0}$,
$K_{S}^{0}\phi$, $K_{S}^{0}\omega$, $K_{S}^{0}\eta$ and
$K_{S}^{0}\eta'$ we require $LR > 0.6$. To give an example of the
performance of this selection, the $LR > 0.4$
requirement keeps 87.5~\% of the $B^{-} \rightarrow D^{0}[\rightarrow
K^-\pi^+]\pi^{-}$signal while removing 73~\% of the continuum
background.

The signal yields are extracted from a fit to the $\Delta E$
distribution in the region $5.27\,\mathrm{~GeV}/c^2 < M_{\rm{bc}} <
5.29\,\mathrm{~GeV}/c^2$. The $B^{-} \rightarrow D^{0}\pi^{-}$ signal
is parameterized as a double Gaussian with peak position and width
floated. On the other hand, we calibrate the shape parameters of the
$B^{-} \rightarrow D^{0}K^{-}$ signal using the $B^{-} \rightarrow
D^{0}\pi^{-}$ data. This accounts for the kinematical shifts and
smearing of the $\Delta E$ peaks caused by the incorrect mass
assignments of prompt hadrons. The peak position and width of the $B^{-}
\rightarrow D^{0}K^{-}$ signal events are determined by fitting the
$B^{-} \rightarrow D^{0}\pi^{-}$ distribution using the kaon mass
hypothesis for the prompt pion, where the relative peak position is
reversed with respect to the origin. The shape parameters for the
feed-across from $B^{-} \rightarrow D^{0}\pi^{-}$ are fixed by the fit
results of the $B^{-} \rightarrow D^{0}\pi^{-}$ enriched sample. The
continuum background is modeled as a first order polynomial function
with parameters determined from the $\Delta E$ distribution for the
events in the sideband region $5.2\,\mathrm{GeV}/c^2 < M_{\rm{bc}} <
5.26\,\mathrm{GeV}/c^2$. Backgrounds from other $B$ decays including
contributions from $B^{-} \rightarrow D^{*0}K^{-}$ and $B^{-}
\rightarrow D^{0}K^{*-}$ are modeled as a smoothed histogram from Monte Carlo simulation. The fit results are shown in Fig.~2.

\begin{table*}[!htb]
\caption{
Signal yields, feed-acrosses and ratios of branching fractions.
The errors on $R^{D}$ are statistical and systematic, respectively.}
\begin{ruledtabular}
\begin{tabular}{lcccc}
Mode & $B^- \rightarrow D \pi^-$ & $B^- \rightarrow D K^- $ & $B
 \rightarrow D \pi^-$& $R^{D} =\frac{{\cal B}(B^- \rightarrow D^0K^-)}{{\cal B}(B^- \rightarrow D^0\pi^-)}$  \\
 &  events               &  events              & feed-across  &  \\ \hline
$ B^- \rightarrow D_{f}h^{-}$ & 6052 $\pm$ 88 & 347.5 $\pm$ 21 & 134.4  $\pm$ 14.7& 0.077 $\pm$ 0.005 $\pm$ 0.006  \\
$ B^- \rightarrow D_{1}h^{-}$ & 683.4 $\pm$ 32.8 & 47.3 $\pm$ 8.9 & 15.6 $\pm$ 6.4 & 0.093 $\pm$ 0.018 $\pm$ 0.008  \\
$ B^- \rightarrow D_{2}h^{-}$ & 648.3 $\pm$ 31.0 & 52.4 $\pm$ 9.0 & 6.3 $\pm$ 5.0 & 0.108 $\pm$ 0.019 $\pm$ 0.007  \\
\end{tabular}
\end{ruledtabular}
\end{table*}
 
\begin{table*}[!htb]
\caption{Yields, partial-rate charge asymmetries and $90~\%$ C.L intervals for asymmetries.}
\begin{ruledtabular} 
\begin{tabular}{lccccc}
Mode & $N(B^{+})$ & $N(B^{-})$ & $\cal{A_{CP}}$ & $90~\%$ C.L \\ \hline
$ B^\pm \rightarrow D_{f}K^{\pm}$ & 165.4 $\pm$ 14.5 & 179.6 $\pm$ 15 & 0.04 $\pm$ 0.06$\pm$0.03 & $-$0.07$<{\cal A}_f<$0.15 \\
$ B^\pm \rightarrow D_{1}K^{\pm}$ & 22.1 $\pm$ 6.1 & 25.0 $\pm$ 6.5 & 0.06 $\pm$ 0.19 $\pm$0.04&  $-$0.26$<{\cal A}_1<$0.38 \\
$ B^\pm \rightarrow D_{2}K^{\pm}$ & 29.9 $\pm$ 6.5 & 20.5 $\pm$ 5.6 & $-$0.19 $\pm$ 0.17$\pm$0.05 &   $-$0.47$<{\cal A}_2<$0.11 \\
\end{tabular}
\end{ruledtabular}
\end{table*}

\begin{figure}[ht]
\begin{center}
 \begin{tabular}{ll}
   \epsfig{file=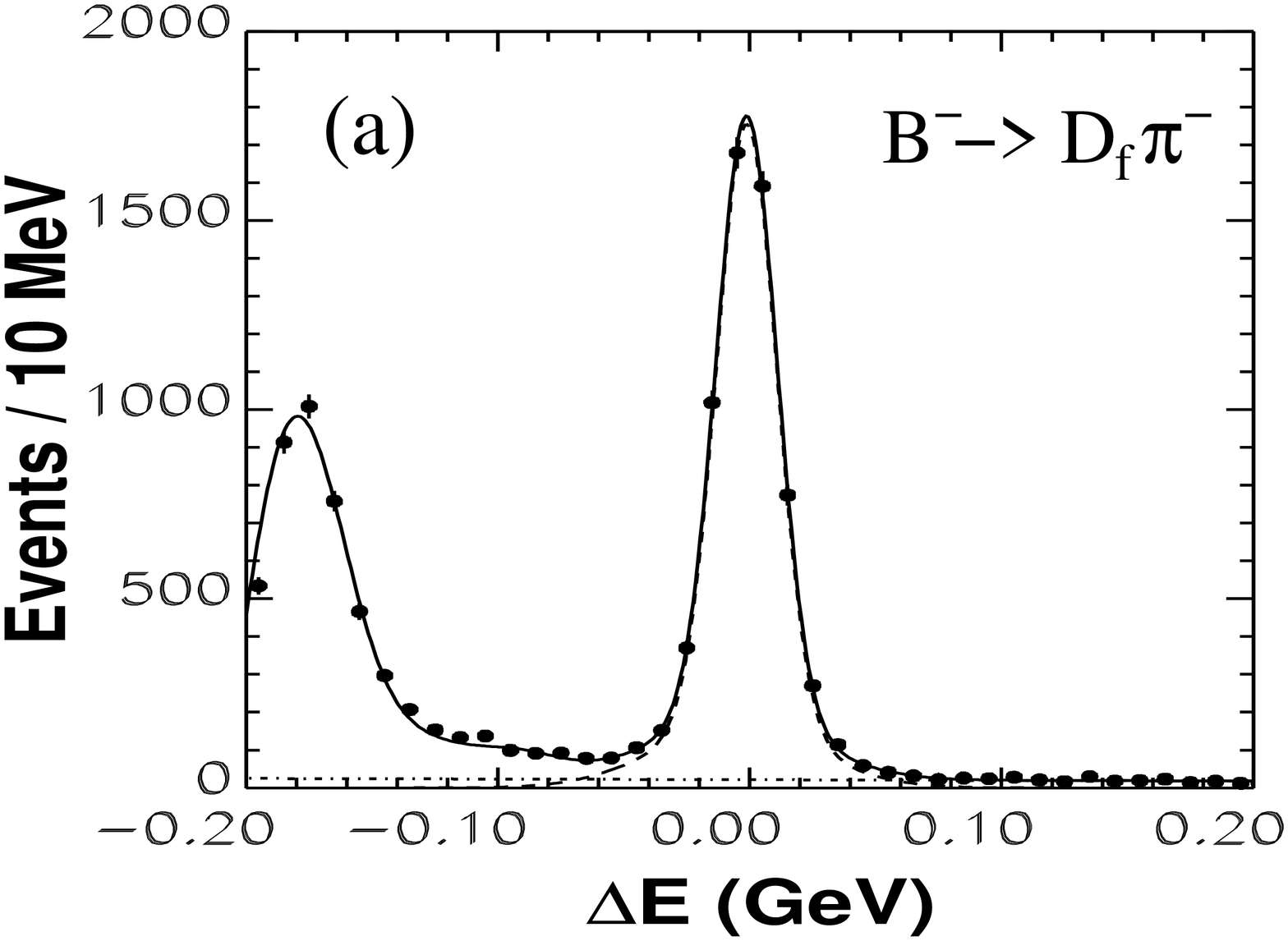,height=5.0cm,width=8.0cm} &
   \epsfig{file=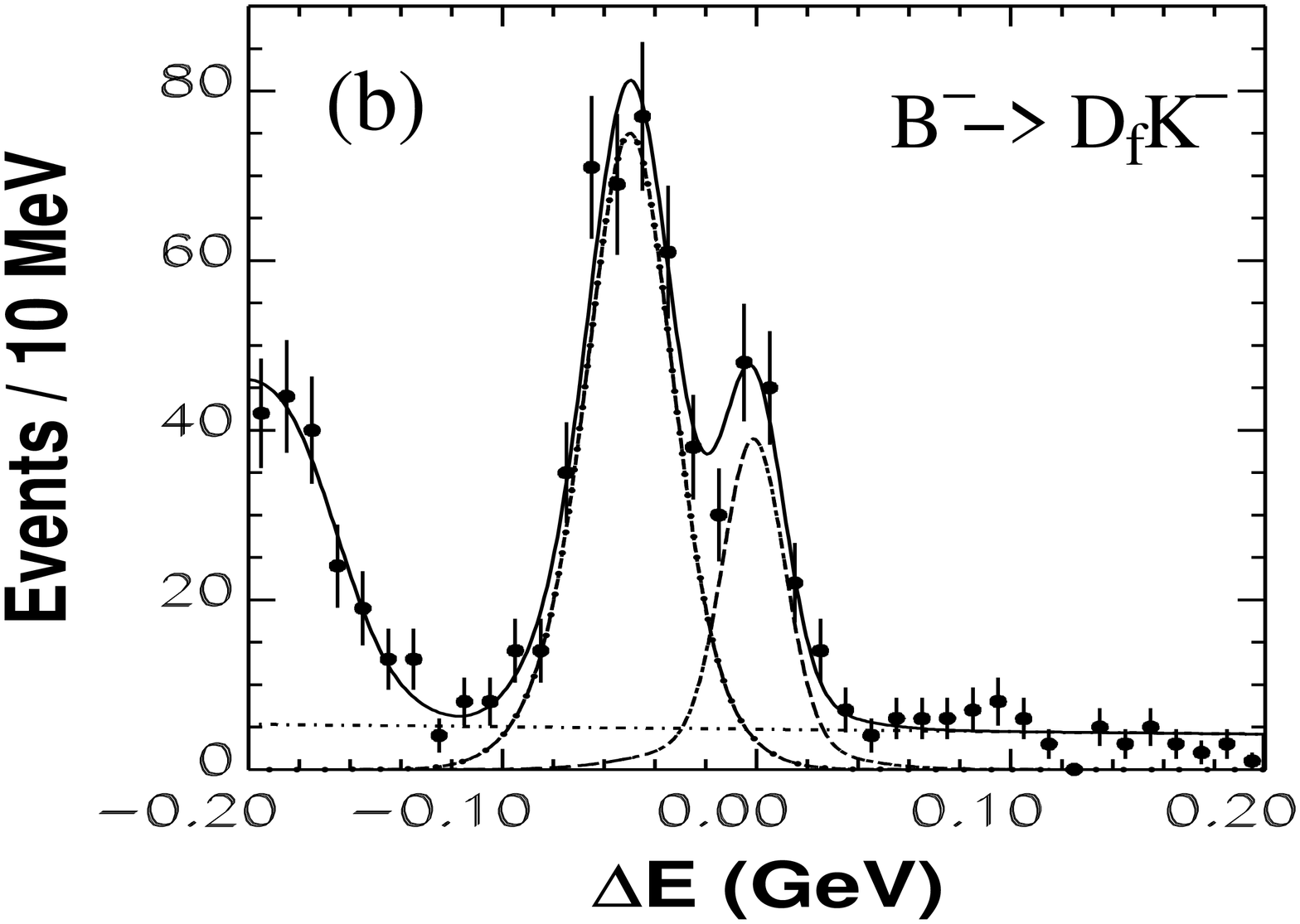,height=5.0cm,width=8.0cm}\\ 
   \epsfig{file=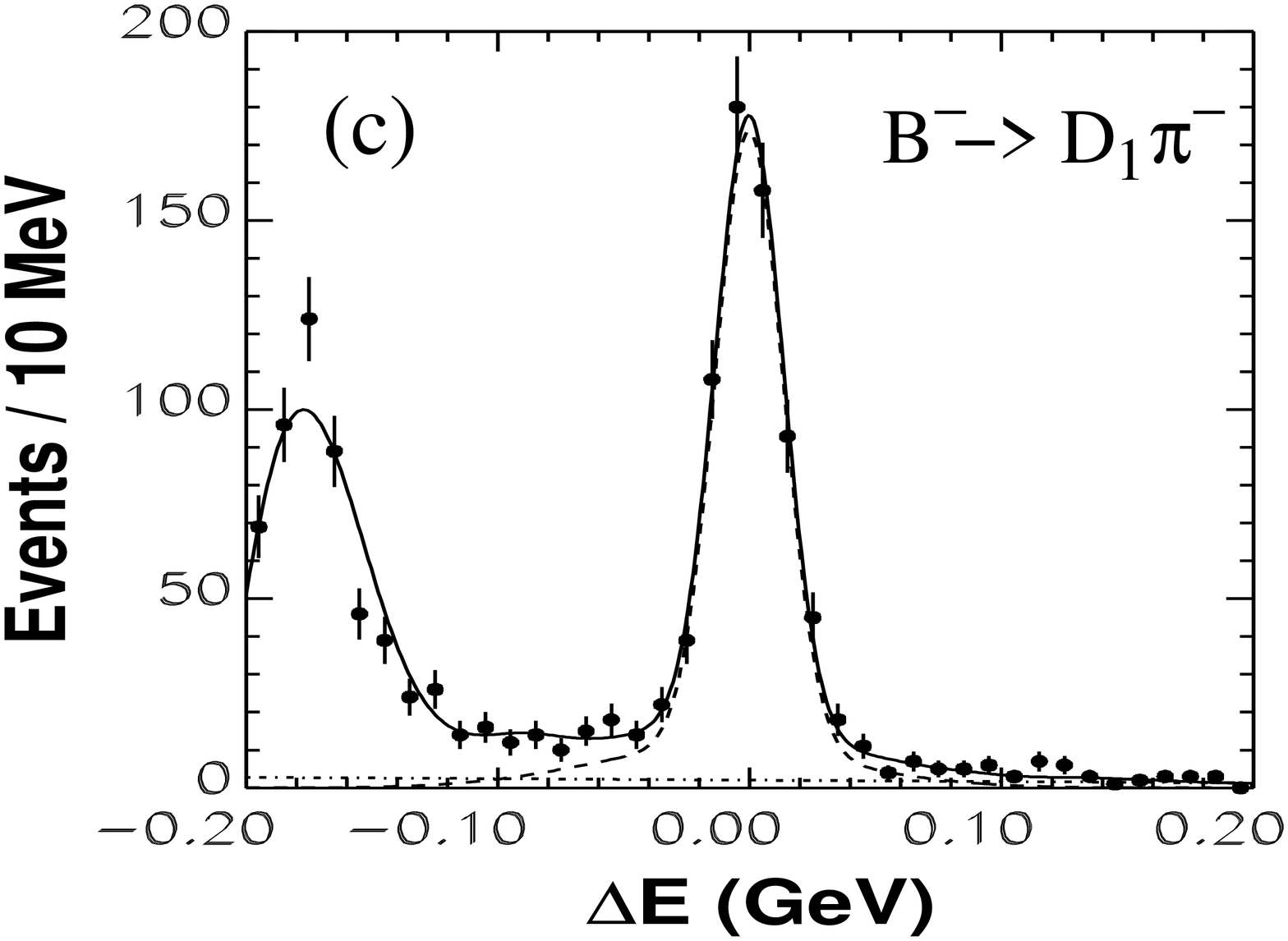,height=5.0cm,width=8.0cm} &
   \epsfig{file=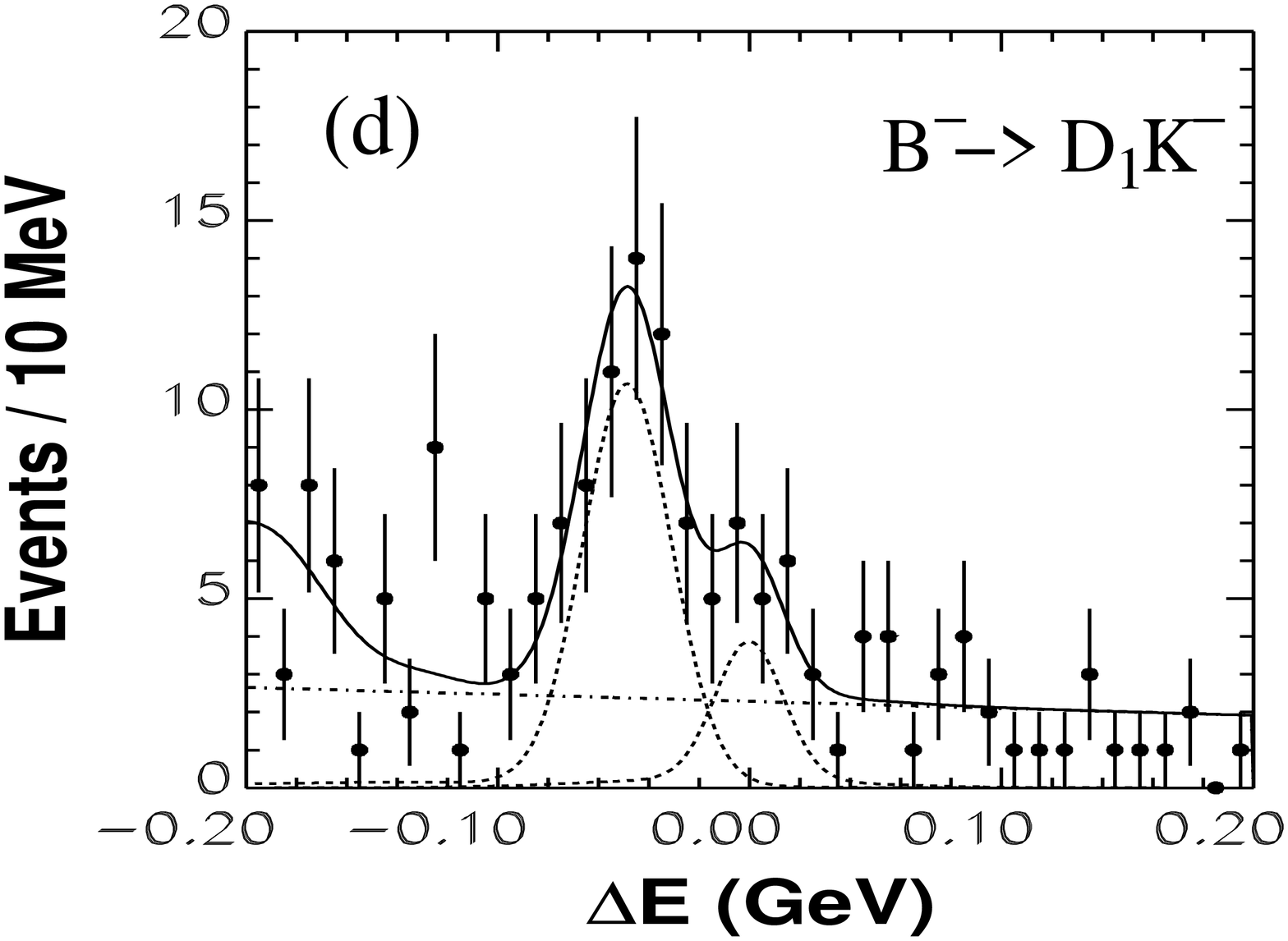,height=5.0cm,width=8.cm} \\
   \epsfig{file=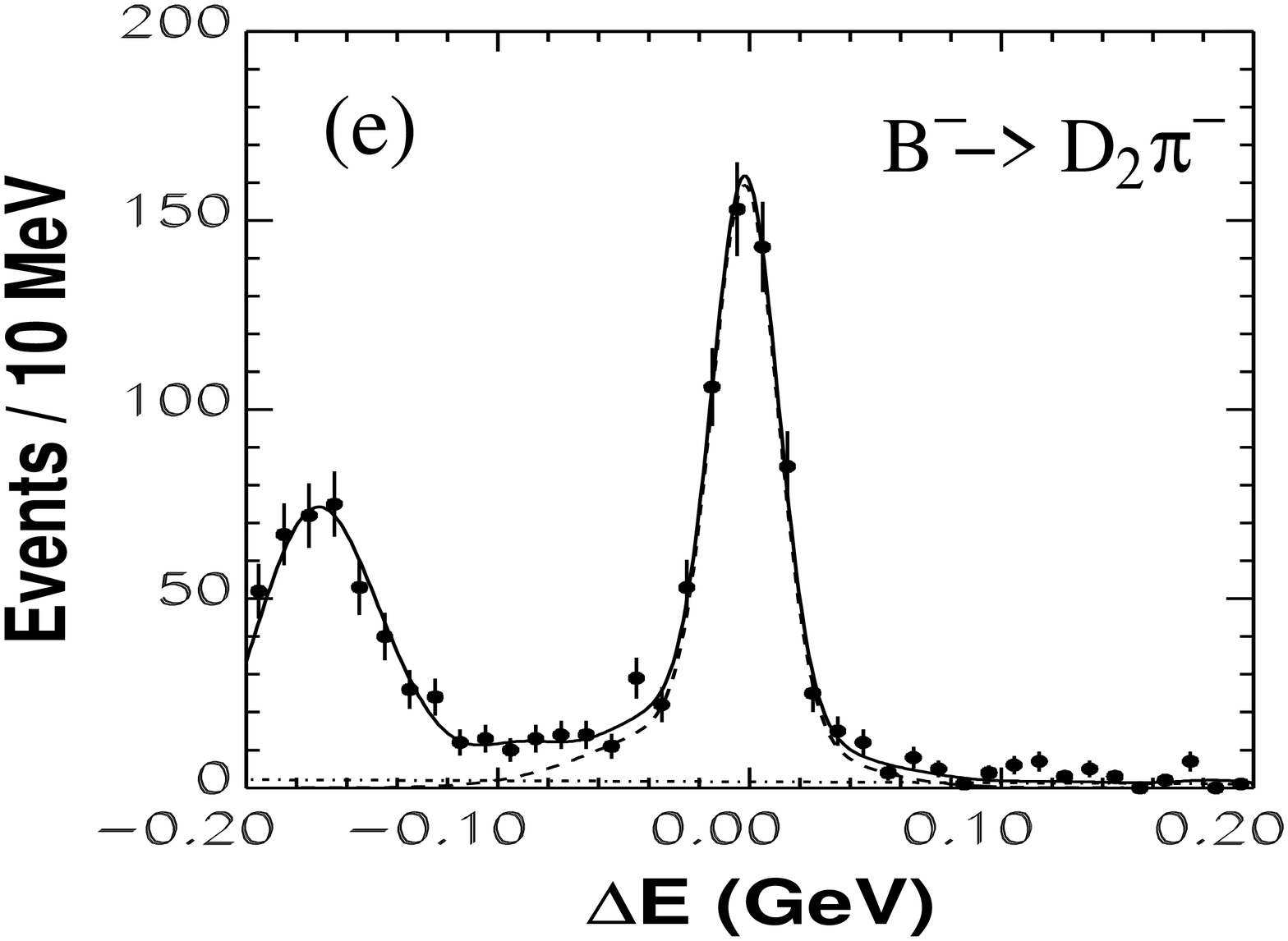,height=5.0cm,width=8.0cm} &
   \epsfig{file=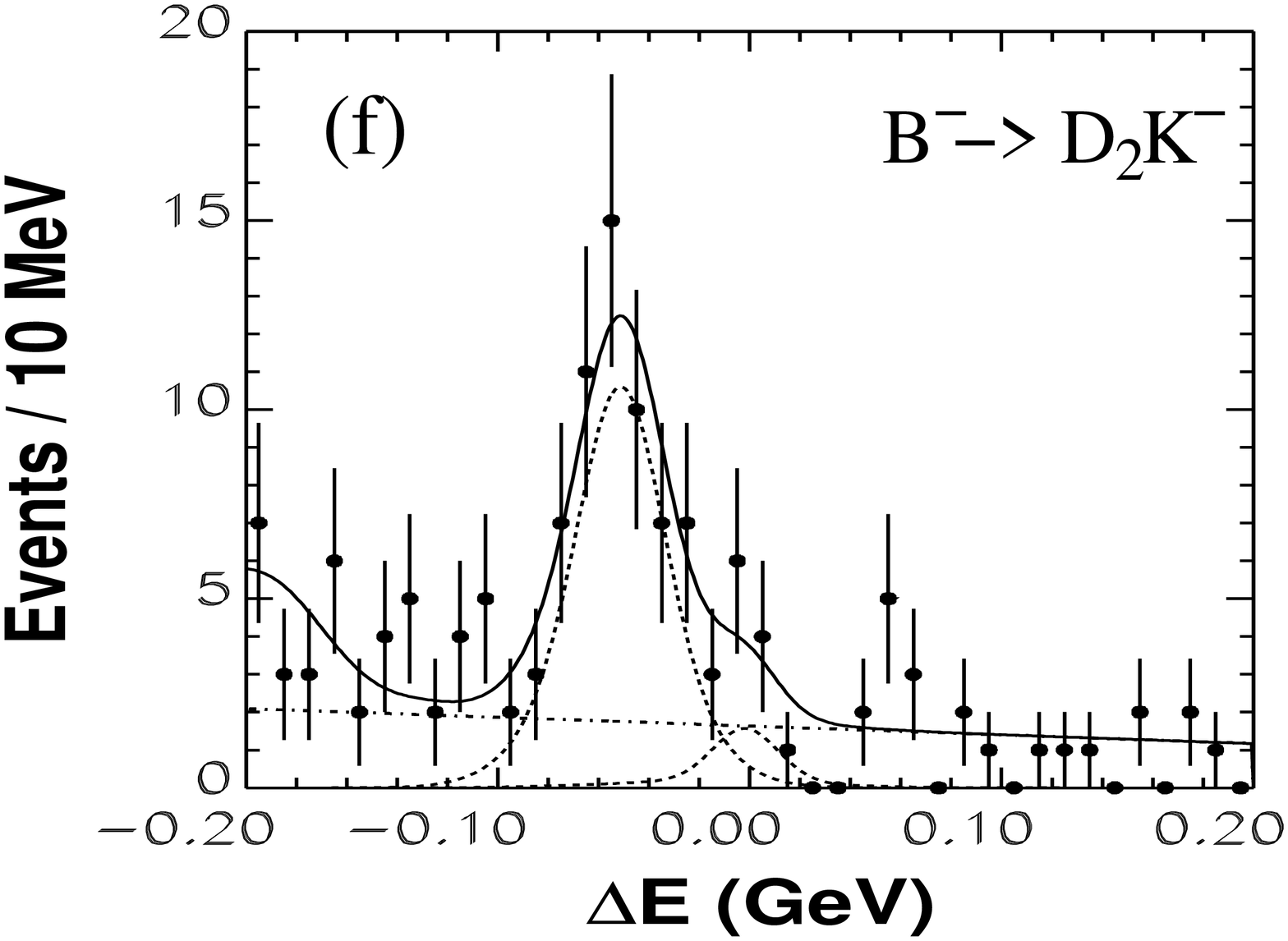,height=5.0cm,width=8.cm} \\
 \end{tabular}
\end{center}
\caption{$\Delta E$ distributions for (a)
$B^{-} \rightarrow D_{f}\pi^{-}$, (b)$B^{-} \rightarrow
D_{f}K^{-}$, (c)$B^{-} \rightarrow D_{1}\pi^{-}$,
(d)$B^{-} \rightarrow D_{1}K^{-}$, (e)$B^{-}
\rightarrow D_{2}\pi^{-}$ and (f)$B^{-} \rightarrow
D_{2}K^{-}$. Points with error bars are the data and the solid lines
show the fit results.}
\end{figure}

The ratios of branching fractions of Cabibbo-suppressed to Cabibbo-favored processes are determined as follows~\cite{PDG},
\begin{eqnarray*}
 R^{D} = \frac{N(B^- \rightarrow D K^-)}{N(B^- \rightarrow D \pi^-)} \times \frac{\eta(B^- \rightarrow D \pi^-)}{\eta(B^- \rightarrow D K^-)} \times \frac{\epsilon(\pi)}{\epsilon(K)},
\end{eqnarray*}
where $N$ is the number of observed events, $\eta$ and $\epsilon$ are
the signal detection and the prompt pion/kaon identification efficiencies,
respectively.  The signal detection efficiencies were determined from a
Monte Carlo simulation e.g. $\eta (B^- \rightarrow D_{f}\pi^-) =$ 44.6~\%
and $\eta (B^- \rightarrow D_{f}K^-) =$ 42.5~\%. The particle
identification efficiencies for the prompt pion and kaon,
$\epsilon(\pi)$ and $\epsilon(K)$, are determined from a kinematically
selected sample of
$D^{*+} \rightarrow D^{0}[\rightarrow K^-\pi^+]\pi^+$ decays, where the $K^-$ and $\pi^+$ mesons from $D^0$
candidates have been selected in the same c.m.\ momentum ~$(2.1 ~{\rm
GeV/}c< p_{c.m.} < 2.5~{\rm GeV/}c)$ and polar angle regions as prompt
hadrons in the $B^- \rightarrow Dh^-$ decay. With our  requirement
$P(K/\pi)> 0.8$, the efficiencies were determined to be $\epsilon(K) = 0.768 \pm 0.001$ and $\epsilon(\pi) = 0.976 \pm 0.001$, and the rate
for misidentification of $\pi$ as $K$ is $0.024 \pm 0.001$. The ratio of $B
\rightarrow D \pi^-$ feed-across  to $B \rightarrow D \pi^-$ signal
is ~2--2.5~\% which is consistent with the measured pion fake
rate. The ratios of Cabibbo-suppressed to Cabibbo-favored decay modes
are shown in Table I. The double ratios are found to be 
\begin{eqnarray*} 
{\cal R}_1 = 1.21 \pm 0.25(stat) \pm 0.14(sys), \\
{\cal R}_2 = 1.41 \pm 0.27(stat) \pm 0.15(sys)~
\end{eqnarray*}
for CP-even and CP-odd eigenstates, respectively. The systematic
 errors in the ratios $R^{D}$ are due to the uncertainty in yield extraction~(3--7~\%) and particle identification~(1~\%). The systematic error in the yield extraction includes uncertainties in the $B\bar{B}$ background and signal shape parametrization. The uncertainty in the $\Delta E$ signal shape parametrization was determined by varying the mean and width of the double Gaussian parameters within their errors. The uncertainty from the slope of the background was determined by changing its value by its error. Both of the resulting changes were included in the systematic error from fitting. Also other backgrounds including rare decays such as $B^{-} \rightarrow K^{-}K^{+}K^{-}$ and $B^{-} \rightarrow K^{-}\pi^{+}\pi^{-}$, which could contribute to the $\Delta E$ signal region, are estimated from the $D^{0}$ sideband data. This uncertainty~(0.5--3~\%) is also included as a source of systematic error. 
 
The asymmetries ${\cal A}_{1,2}$ are evaluated using signal yields
obtained from separate fits to the $B^{+}$ and $B^{-}$ samples shown
in Fig.~3. The results are given in Table II. We find 
\begin{eqnarray*} 
{\cal A}_1 &=& 0.06 \pm 0.19(stat) \pm 0.04(sys), \\
{\cal A}_2 &=& -0.19 \pm 0.17(stat) \pm 0.05(sys)
\end{eqnarray*}
where the systematic uncertainty is from the intrinsic detector
charge asymmetry~(3.2~\%), the $B^-$ and $B^+$ yield extractions~(2.4--3.3~\%), and the asymmetry in particle identification
efficiency of prompt kaons~(1~\%). The intrinsic detector charge
asymmetry is calculated from the $B^{-} \rightarrow
D^{0}[\rightarrow K^-\pi^+]\pi^{-}$ sample. The systematic error from yield extraction
is calculated by changing the fitting parameters by $\pm 1\sigma$. 

In summary, using $78~{\rm fb}^{-1}$ of data collected with the Belle
detector, we report measurements of the decays $B^{-} \rightarrow
D_{CP}K^{-}$, where $D_{CP}$ are the neutral $D$ meson CP
eigenstates. These supersede the results reported in~\cite{belledk,sugi}. The
ratios of the branching fractions $R^{D_{1,2}}$ for the decays $B^{-}
\rightarrow D_{CP}K^{-}$ and $B^{-} \rightarrow D_{CP}\pi^{-}$ are
consistent with those for the flavor specific decay within errors. The
measured partial-rate charge asymmetries ${\cal{A}}_{1,2}$ are
consistent with zero. 

\begin{figure}[ht]
\begin{center}
 \begin{tabular}{ll}
   \epsfig{file=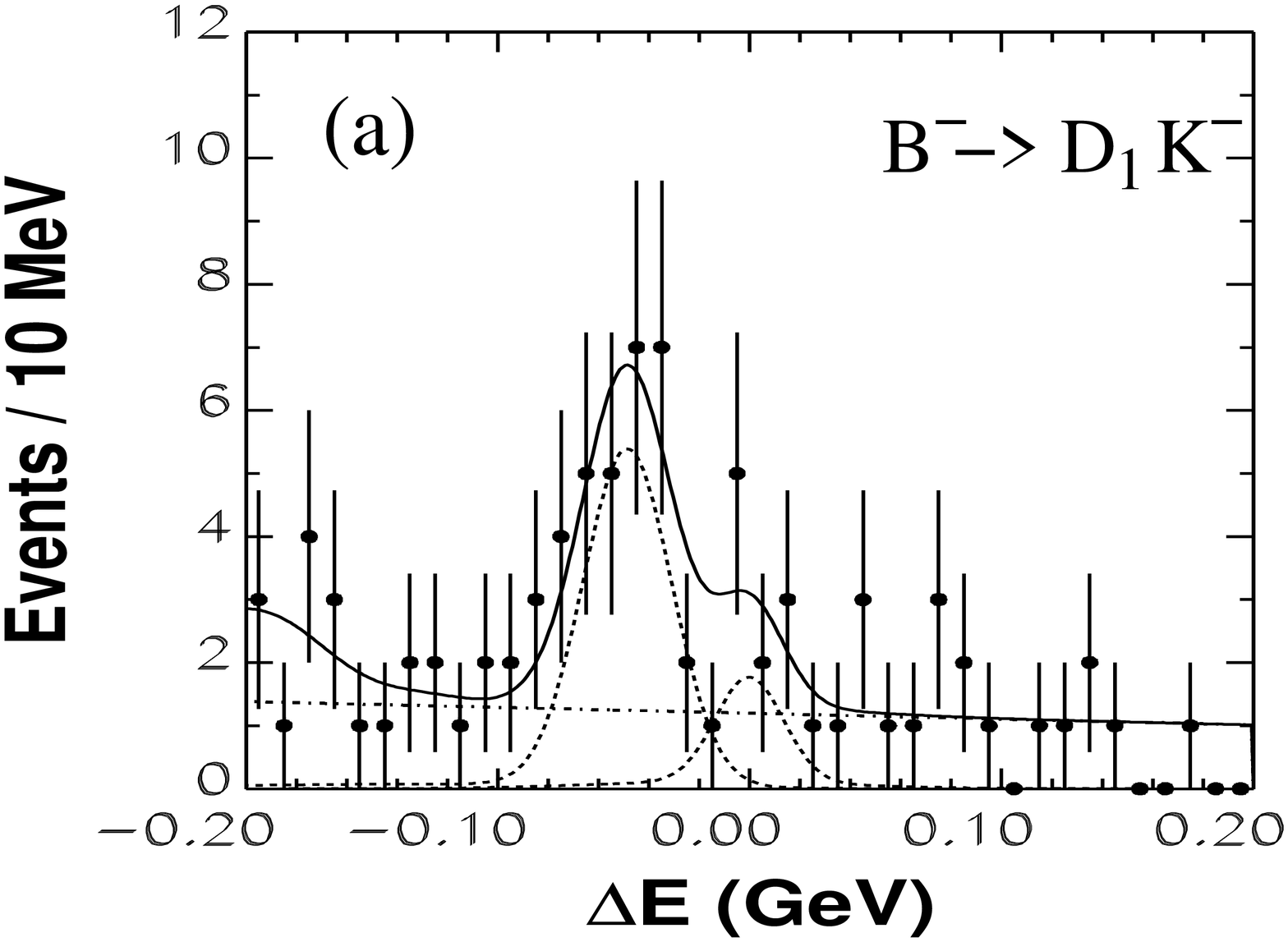,height=5.0cm,width=8.0cm} &
   \epsfig{file=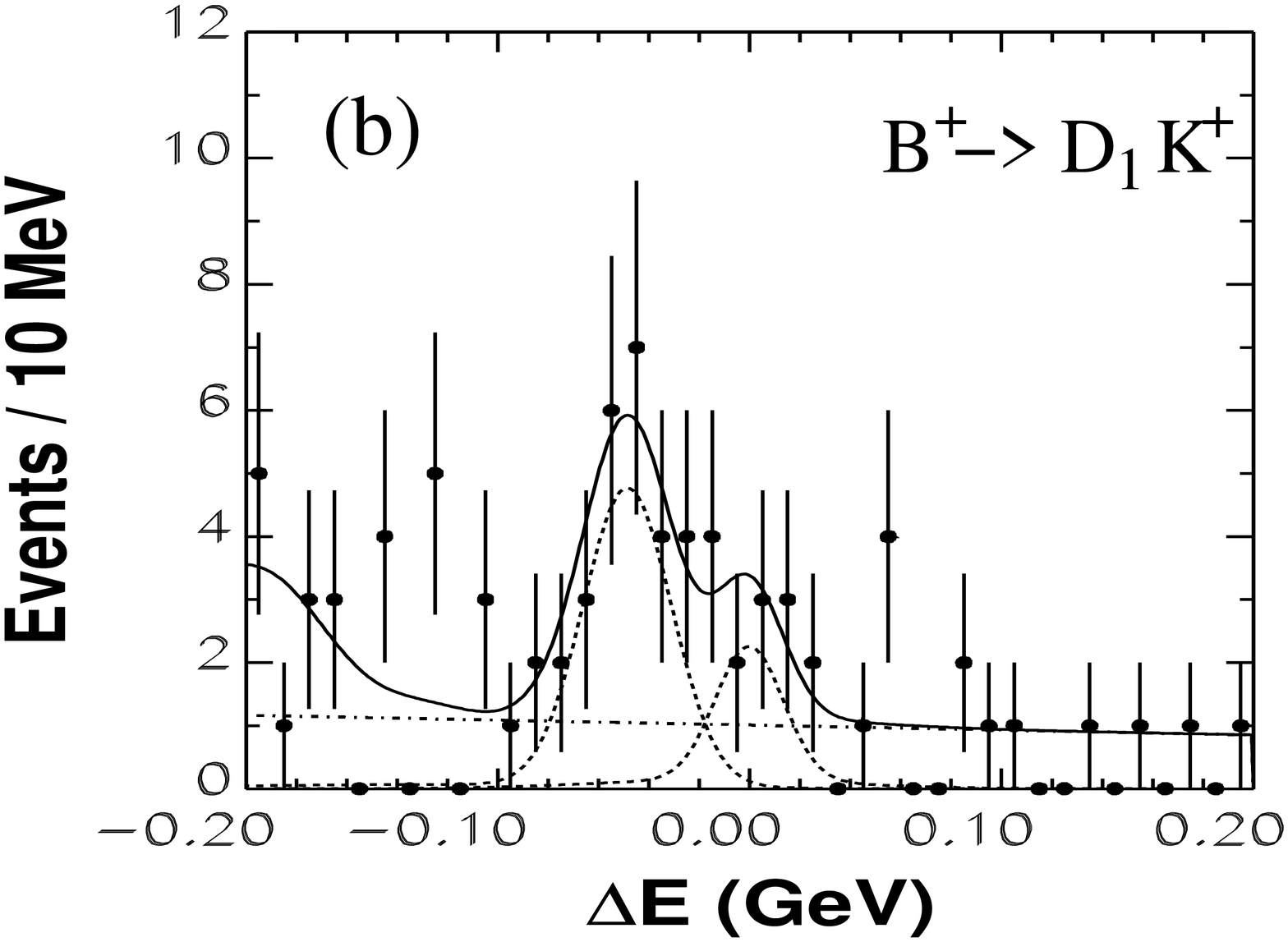,height=5.0cm,width=8.cm} \\
   \epsfig{file=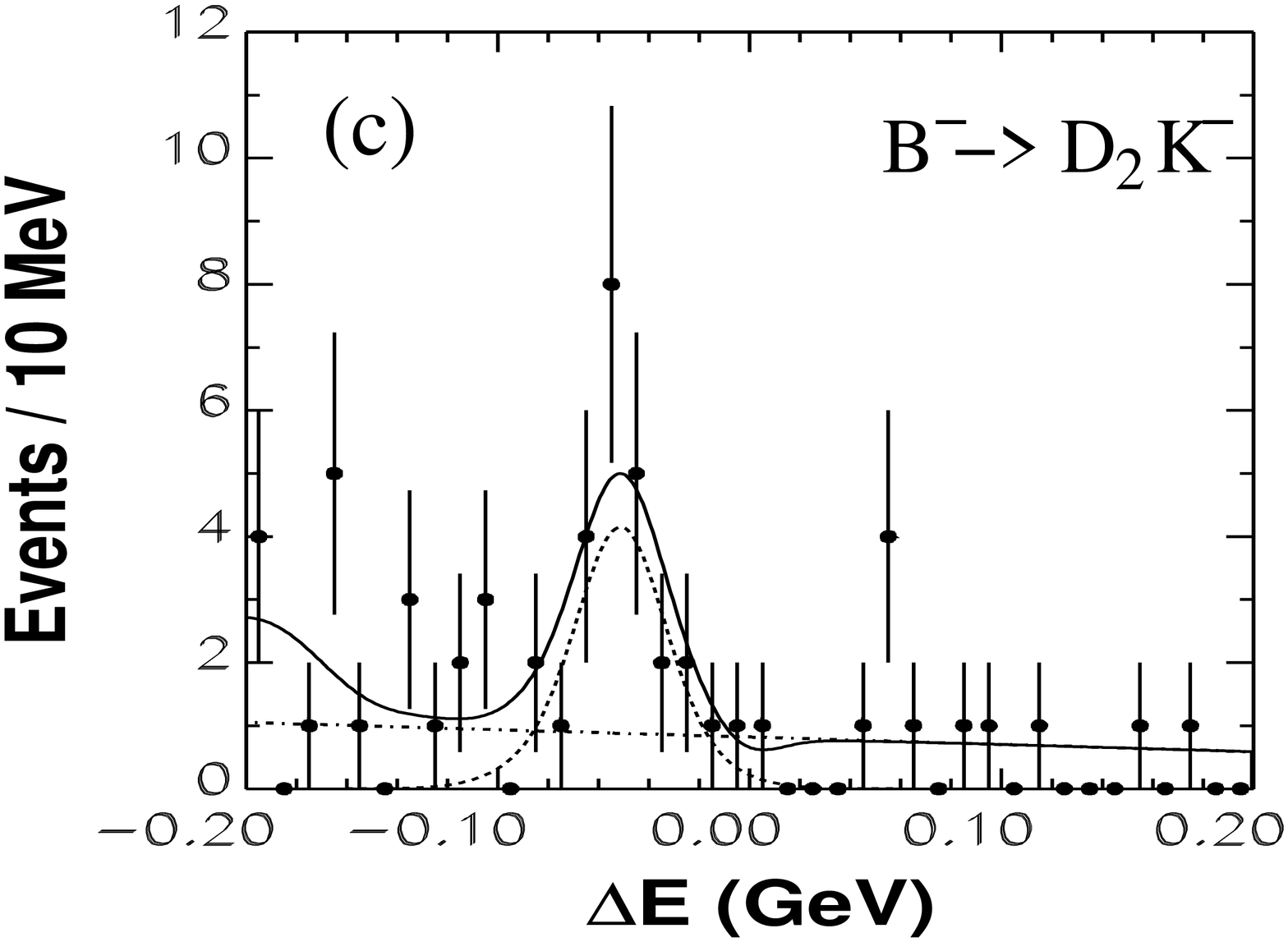,height=5.0cm,width=8.0cm} &
   \epsfig{file=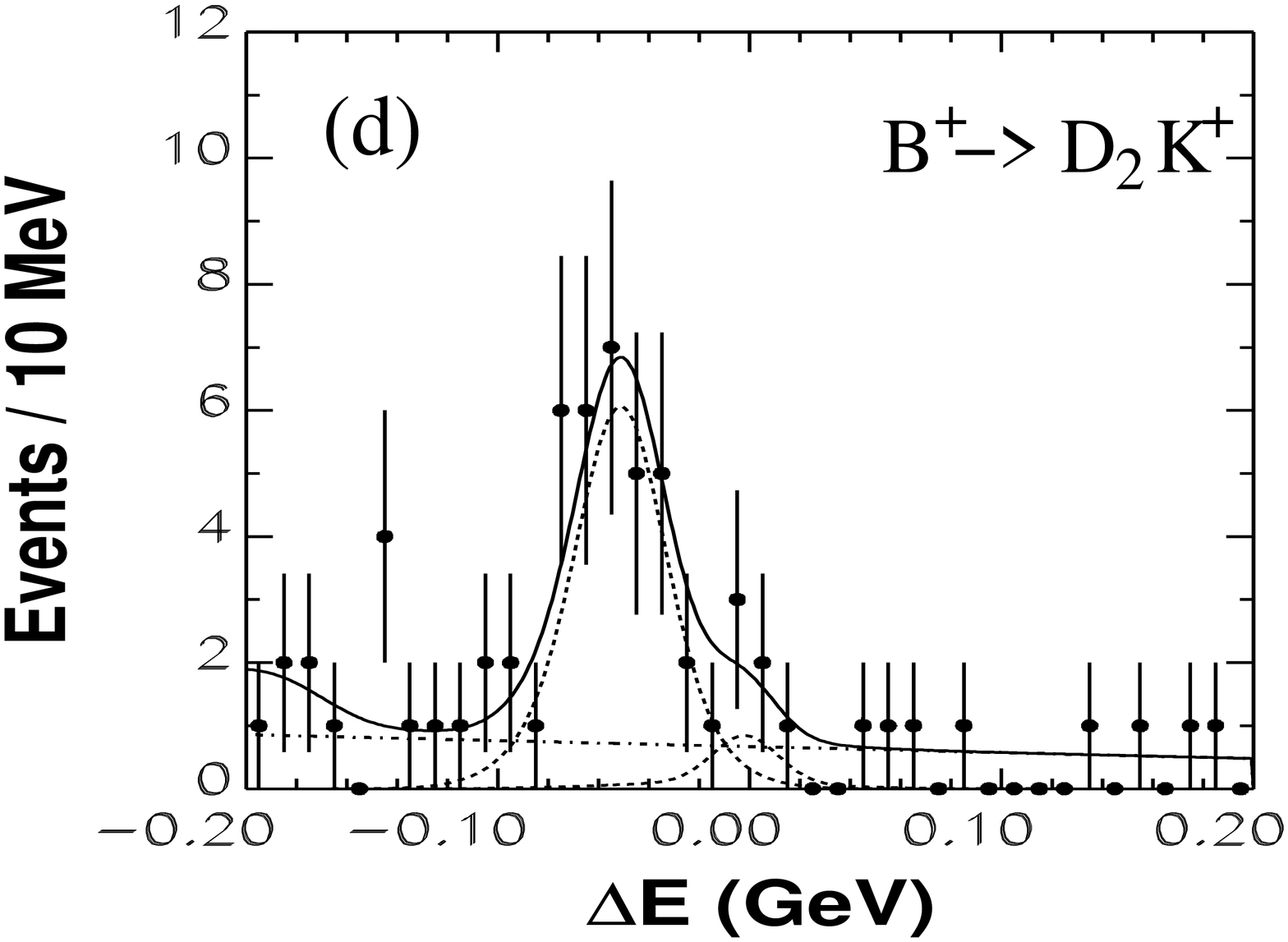,height=5.0cm,width=8.cm} \\
 \end{tabular}
\end{center}
\caption{$\Delta E$ distributions for the charge conjugate modes (a)
$B^{-} \rightarrow D_{1}K^{-}$, (b)$B^{+} \rightarrow
D_{1}K^{+}$, (c)$B^{-} \rightarrow D_{2}K^{-}$,
(d)$B^{+} \rightarrow D_{2}K^{+}$.}
\end{figure}

\section*{Acknowledgments}
We wish to thank the KEKB accelerator group for the excellent
operation of the KEKB accelerator.
We acknowledge support from the Ministry of Education,
Culture, Sports, Science, and Technology of Japan
and the Japan Society for the Promotion of Science;
the Australian Research Council
and the Australian Department of Industry, Science and Resources;
the National Science Foundation of China under contract No.~10175071;
the Department of Science and Technology of India;
the BK21 program of the Ministry of Education of Korea
and the CHEP SRC program of the Korea Science and Engineering Foundation;
the Polish State Committee for Scientific Research
under contract No.~2P03B 17017;
the Ministry of Science and Technology of the Russian Federation;
the Ministry of Education, Science and Sport of the Republic of Slovenia;
the National Science Council and the Ministry of Education of Taiwan;
and the U.S.\ Department of Energy.


\begin{thebibliography}{99}
\bibitem{gamma} Another naming convention, 
$\gamma$ ($= \phi_3$), is
also used in the literature.
\bibitem{CC}
Hereafter, the inclusion of the charge conjugate mode
decay is implied unless otherwise stated.
\end{thebibliography}
\end{document}